# Histological and Histochemical Studies on the Esophagus, Stomach and Small Intestines of *Varanus niloticus*

## Ahmed YA[1], El-Hafez AAE[2], Zayed AE[2]

[1] Faculty of Veterinary Science, South Valley University, Qena, Egypt.
[2] Faculty of Veterinary Medicine, Assiut University, Assiut, Egypt.
*Email:* ya_abd@yahoo.com

## Abstract

Information on the digestive system of the reptiles is based on relatively few studies on some of the now present 7500 reptilian species.  Yet, the gap between our understanding of the major similarities and / or differences between the mammalian and reptilian digestive system does not seem satisfiable.  The aim of the current study was to investigate the morphological structure of one of the most common reptilian species in Egypt, *Varanus niloticus* or Nile monitor.  Specimens for histological examination were collected from the esophageus, stomach and small intestine of the Nile monitor and pro-cessed for paraffin embedding.  Sections were stained with haemat-oxylin and eosin for general morpho-logy.  Periodic Acid Schiff's (PAS) and Alcian Blue (AB) staining methods were applied to detect the different types of the mucous contents of the gastro-intestinal tract.  Some paraffin sections were stained with Grimelius silver impregnation technique for localization of the enteroendo-crine cells.

The folded esophageal mucosa had ciliated columnar epithelium with muc-ous secreting goblet cells, which stained positive with PAS and AB.

The esophageal mucosa was folded and the lining epithelium was ciliated columnar epithelium with mucous se-creting goblet cells, which stained posi-tive with PAS and AB.

The stomach was divided into fundic and pyloric regions.  The mucosa was thrown into gastric pits, into which the gastric glands opened.  The surface epithelium was mucous secreting co-lumnar cells and stained positive with PAS but negative with AB.  The fundic gland was made by oxynticopeptic cells and few mucous cells, while the entire pyloric gland was made by mucous cells stained positive for both PAS and AB.  The small intestine showed many villi but occasional poorly developed intestinal crypts were.  The intestinal mucosa was lined with absorptive co-lumnar epithelium with goblet cells, which stained positive with PAS and AB.  Enteroendocrine cells were of dif-ferent shapes; rounded, spindle, oval or pyramidal and were localized in the sur-face epithelium of the esophagus and small intestine, and among the cells of the gastric glands.

This is the first report about the histolo-gy of the alimentary tract of the *Varanus niloticus*. Further studies are required for investigation the possibility of using this animal as a model for studying the regulation of the digestive processes as well as to understand the theory of de-velopment.

## Introduction

It has been suggested that reptiles could serve as a model" Reptiles have been suggested to be a future useful model for studying the physiological regulation of the digestive process as they have well responses to feeding even more than other commonly used experimental mammals such as mice, rats, rabbit and pigs (Secor and Di-amond, 1998). Reptiles include as many as 7500 different species, most known are; alligators, turtles, tortoises, lizards and snakes (Elliott, 2007).

One of these reptiles is the *Varanus niloticus* or the Nile monitor, which is the biggest African lizard and one of the most voracious predators (Capula, 1990). It lives along the Nile; it has been found mainly in swamps and lakes (Smith *et al.*, 2008). Nile monitor has a powerful body with an elongated snake-like head, sharp claws, and a long and strong tail, which is used to defend itself when threatened (Alden *et al.*, 2005). This species of reptiles used to eat any thing it can overpower or find as a car-rion such as arthropods, frogs, fish, birds, small mammals and other smaller reptiles as well as the eggs of the Nile crocodiles (Capula, 1990; Smith *et al.*, 2008 ). Depending upon its powerful tail, the Nile monitor is an excellent swimmer and it can stay under the wa-ter for up to one hour (Capula, 1990), how-ever, it is not only an aquatic ani-mal, but also walks and sometimes climbs trees to feed, bask or sleep (Al-den *et al.*, 2005). Mating of the Nile monitors occurs after rainy seasons, and the female give about 60 eggs that are incubated for a period of 6-9 months before hatching, and the offspring reach their maturity after about 4 years (Ca-pula, 1990).

The digestive system of the reptiles contains all the structures present in other higher vertebrates, from the oral cavity to the cloaca. The oral cavity is lined by mucous membrane made by non-keratinized stratified squamous epithetlium with salivary glands distrib-uted in the submucosa (Putterill and Soley, 2003).

The alimentary tract of reptiles is similar to higher vertebrates with some excep-tions. The esophagus shows adaptive modifications from group to group. In turtles, the esophagus has heavily kera-tinized papillae that protect the mucosa from abrasive diet such as speculated sponges and jellyfish, and also may act as filtering devices. In lizards, it is formed of folds lined by ciliated columnar epithelium with goblet cells. Some snakes have mucous glands along their submucosa (Elliott, 2007). The muscularis mucosa of the esophagus is absent in many species of reptiles but may be found in some species of turtles (Elliott, 2007).

In reptiles the stomach varies in shape (Elliott, 2007). The stomach of turtles has greater and lesser curvatures, crocodiles have a saccular stomach, lizards have an ovoid one, while the stomach of snakes is elongated in shape (Madrid *et al.*, 1989). The mucosa of the stomach is folded and the extend of folding varies among different groups of reptiles. Two parts of the reptilian stomach are usually distinguishable, the fundus and the pylorus (Luppa, 1977). The mucosa of the fundic region consists of rows of branched tubular glands. Each gland consists of a gastric pit and glandular body. Mucous neck cells similar to that of mammals may be present in some species such as snakes or absent in others such as turtles (Elliott, 2007). The glandular portion contains either one type or two types of cells; dark serous (oxynticopeptic) and clear mucous cells. The ratio of the two cells also may be different from species to another (Suganuma *et al.*, 1981). The dark cells functions as both parietal and chief cells of mammals. Additional cells, enteroendocrine cells are scat-tered throughout the digestive tract. These cells contain mem-braned vesicles of peptide hormones and amines (D'Este *et al.*, 1995), and function similar to that of mammals (Krause *et al.*, 1985).

The reptilian small intestine may be highly convoluted as in turtles or relatively straight as in snakes (Parsons and Cameron, 1977). The intestinal surface is varies among different groups; it may be thrown into longitudinal and transverse folds, have a zig-zag pattern or netlike or honeycomb appearance, or consists of fine striations (Parsons and Cameron, 1977). The mucosa is organized into villi with poorly developed intestinal crypts in most species or crypts maybe absent (Luppa, 1 977). The lining epithelium of the small intestine is formed of absorptive columnar cells and goblet cells. Enteroendocrine cells can be identified using Grimelius silver stain (Burrell *et al.*, 1992). Beyond the duodenum, it is difficult to distinguish between different parts of the small intestine.

## Materials and Methods

### *Sample Collection and Paraffin Embedding*

An adult female Nile monitor, 1.5 meters long was collected from a lake in Qena close to the Nile River. The animal was killed and the alimentary tract was removed, opened, cleaned and samples were taken from the esophagus, stomach (fundic and pyloric regions) and different parts of the small intestine. Samples were thoroughly washed in phosphate buffered saline and rapidly immersed in either 10% paraformaldehyde for at least 3 days or Bouin's fixative for 30 minutes. The samples were dehydrated in an ascending graded ethanol series (70, 90, 95 and 100%) for 1 hour each, and then cleared in 3 changes of, xylene for 24 hours. The cleared specimens were then embedded in melted paraffin wax.

### Histological staining

Paraffin embedded specimens were sectioned at 6 μm thickness using a rotary microtome. Sections were de-paraffinized and stained with haemato-xylin and eosin for general morphology. For detection of mucous in the alimentary tract of the *Varanus niloticus* , sections were stained with periodic acid Schiff, alcian blue (PAS-AB; pH 2.5) to demonstrate the full comple-ment of tissue proteoglycans; alcian blue (AB) stain specifically acidic mucins while PAS stain neutral one. Grimelius silver impregnation tech-nique was used for detection of entero-endocrine (argyro-philic) cells.

## Results

### Histological Structure of the Mucosa and characterization of the Mucinous Contents of the Alimentary Tract of the Varanus niloticus

The alimentary tract of the *Varanus niloticus* revealed the typical layers seen in higher vertebrates; mucosa, submucosa, tunica muscularis and serosa (Figs. 1A-D). The esophagus was a short tube and its mucosa was folded (Fig. 1A). The esophageal mucosa contained extensive primary and secondary folds, and the lining epithelium consisted of ciliated colum-nar epithelium and mucous secreting goblet cells (Figs. 2A, B), which stained positive with both PAS and AB (Fig. 3A). The stomach of *Varanus niloticus* was sacu-lar in shape and its mucosa thrown into longitudinal folds. Histo-logically, the stomach divided into two distinct parts; the fundus or body and the pylorus (Figs. 1B, C). The stomach was lined through its length with mucous-secreting columnar epithelium that showed numerous invaginations, gastric pits, which led to glandular structures, the gastric glands (Figs. 2C, D). The gastric surface epithelium showed the same histological features in both fundic and pyloric regions, and exhibited strong staining with PAS but did not react with AB (Fig. 3B). In the fundic region, the epithelium forms deep glands, fundic glands. The fundic glands are branched tubular and formed primarily of dark oxynticopeptic cells with some clear mucous cells located between the gastric pits and fundic glands (Fig. 2D). We observed that the oxynticopeptic cells showed different gradient of staining; some cells were deeply stained, while, some other cells appeared granulated or nearly empty. The oxynticopeptic cells were cuboidal or pyramidal in shape and stained deeply acidophilic but they showed dif-ferent gradient of staining; some cells were deeply stained, while, some other cells appeared granulated or nearly empty.. Oxynticopeptic cells stained negative with both PAS and AB. The mucous cells appeared clear or with foamy cytoplasm and stained positive with PAS and AB. The pyloric glands were simple tubular, shorter than those in the fundic region and consisted only of numerous mucosecreting cells, which showed positive reaction with PAS and AB (Fig. 3C). Mitotic figures were often seen among the mucous cells of the gastric glands in both parts. Lamina propria of fundic and pyloric region was loose connective tissue with many lymphocytes and eosinophils. The submucosa was connective tissue layer with blood vessels. Underneath the submucosa, the tunica muscularis consisted of inner circular and outer longitudinal smooth muscle cells. The pyloric region showed well developed layers of smooth muscles (Fig. 1C).

The structure of the small intestine appeared uniform throughout its entire length. The intestinal mucosa of the small intestine folded into villi with no or few intestinal crypts (Fig. 2D). It was difficult to distinguish between different parts of the small intestine. Two basic types of cells are present in the intestinal lining epithelium; columnar absorptive cells and goblet cells that secret both types of the mucinous substances as indicated by positive reaction to both PAS and AB. The epithelium covering (epithelium covers the villi surface not lines) the villi was simple columnar with goblet cells. No Paneth cells were seen. The lamina propria was rich in lymphocytic infil-trations and many eosinophils. A not well developed layer of circular muscularis mucosa was present. The submucosa was extremely narrow, and the circular and longitudinal layers of the tunica muscularis contained distinct layers of dense fibrous connective tissue (Fig. 2D).

***Localization of Enteroendocrine Cells in the Alimentary Tract of the Varanus niloticus***

Grimelius silver stain was applied on the paraffin sections to localize the enteroendocrine cells. Enteroendocrine cells were oval or spindle in shape and found in all parts of the alimentary tract examined (Fig. 4A, B). They were distributed in the surface epithelium of the esophagus and intestine (Fig. 4A), where they made contact with the lumen via cytoplasmic processes. In the stomach, they were very few in the surface epithelium, but they were seen in abundance within the pyloric glands. In stomach, the enteroendocrine cells appeared pyramidal, oval or spindle in shape and made contact with the basement membrane between mucous cells.

## Discussion

The aim of this study was to document the histology of the alimentary tract of a common Egyptian reptilian species, *Varanus niloticus* or the Nile monitor. Classic special histologic techniques were used.

Similar components as in higher vertebrates with few exceptions were found in the alimentary tract of this species. The lining epithelium of the esophagus was ciliated columnar cells with mucous secreting cells. Although, the lining epithelium of the esophagus of some reptiles is similar to that of mammals; stratified squamous epithet-lium but it is also known that some species are different and lined with columnar epithelium (Elliott, 2007) and that is in agreement with the result reported here. The mucous secreting cells stained positive with PAS and AB, suggesting that they secrete both neutral and acidic muco-substances. The mucous, especially the acidic one, which increases the viscosity of the mucous contents may be important for lubrication the mucosa and allowing the passage of the large food particles, in addition to immobilization of overcome hunted food such as arthropods and mice. Also, the mucous is important to protect the mucosa from the sharp objects may reach the esophagus such as spiny fish. The presence of both neutral and acidic mucin in the lining epithelium of esophagus is not unique for the *Varanus niloticus*, but it was seen in other reptiles (Elliott, 2007) and fish species (Parillo *et al.*, 2004), crocodiles (Uriona *et al.*, 2005) and in many birds (Selvan *et al.*, 2008). Unlike the mammalian esophagus (Eurell and Frappier, 2006), the muscular layers of the esophagus of *Varanus niloticus* is entirely smooth. Therefore, it may allow the esophagus to expand and to tolerate the passage of the large food particles such as mice or fish.

Similar to other reptiles; King's skink (Arena *et al.*, 1990) and lizard (Ferri *et al.*, 1999), the fundic glands of the stomach of the *Varanus niloticus* contained only one cell type, oxyntico-peptic cells. In some other reptiles such as crocodiles it was reported that the morphology of the oxyntico-peptic cells changes from the proximal to the distal mucosa of the stomach. At the proxymal portion of the stomach, the oxynticopeptic cells have characteristic features of protein synthesizing cells; well developed rough endoplasmic reticulum and many granules, similar to pepsinogen-secreting chief cells of mammals.

At the distal portion of the stomach, oxynticopeptic cells contain few zymogen granules, numerous mito-chondria and a well developed tubule-vesicular system typical of the mammalian parietal acid-secreting cells (Giraud *et al.*, 1979; Ferri *et al.*, 1999). Thus, it is not surprising that oxyn-ticopeptic may elaborate both hydro-chloric acid pepsin0gen synthesized by the parietal and chief cells in mammalian stomach (Eurell and Frappier, 2006). It is not unique to reptiles that one cell type performs both functions of parietal and chief cells of mammals. It is also the case in fowl (Selvan *et al.*, 2008) and some fish (Ota *et al.*, 1998). We observed that the oxyntico-peptic cells showed different gradient of staining; some cells were deeply stained, while, some other cells appeared granulated or nearly empty. Variation of staining of the oxyntico-peptic cells may due to a secretion gradient of hydrochloric acid and/ or pepsingen, as hydrochloric acid is essential for the activation of pepsinogen.

The superficial epithelium of the stomach showed positive reaction with PAS but not with AB. It is likely that, this cell layer is responsible for secretion of neutral mucosubstances, and that in similar to other reptilian species (Arena *et al.*, 1990; Ferri and Liquori, 1992; Ferri and Liquori, 1994; Ferri *et al.*, 1999). While, we found mucous-secreting cells in the upper part of the fundic glands and in the whole pyloric glands elaborate both acidic and neutral mucosubstances as they stained positive for both PAS and AB and that similar to mucous neck cells of the mamma-lian such as guinea pig, gastric glands (Sato and Spicer, 1980), but different from some lizards and crocodiles where the mucous cells secrete only neutral mucous (Madrid *et al.*, 1989) and other reptiles such as some species of snakes in which the mucous cells elaborates strong acidic mucosubstances (Suganuma *et al.*, 1981). The mucous is important to protect the gastric mucosa from the effect of hydrochloric acid and pepsin in the fundic region, but in the pylorus it may be primarily important as lubricant. Mitotic figures were seen among the glandular mucous cells. These cells may act as stem cells for different cell type regeneration in the glandular epithelium.

The structure of the small intestine appeared uniform throughout its entire length. Two basic types of cells are present in the intestinal lining epithelium; columnar absorptive cells and goblet cells that secret both types of the mucinous substances as indicated by positive reaction to both PAS and AB." The intestinal crypts were less developed. In mammals, proliferative activity is confined to the intestinal crypts where undifferentiated stem cells give rise to new cells that continually replace those are lost at the tips of each villus (Gilbert, 1997; Junqueira *et al.*, 1998). It was not clear using the method reported in this study to know where the stem cells are located and how the lining epithelium is replaced in the intestinal tract of the *Varanus niloticus*. Below the epithelium is the lamina propria consisting of blood vessels and connective tissue. The muscularis mucosa is very thin and consists of a single layer of smooth muscle fibers. The submucosa is below the muscularis mucosa, followed by the tunica muscularis and then serosa. The small intestine histology was generally typical to crocodiles and lizard (Holmberg *et al.*, 2002; Elliott, 2007).

Enteroendocrine cells are another cell type found in the alimentary tract of the Varanus *niloticus* and were scattered among the surface epithelium of esophageal folds and intestinal villi. In stomach, they were located among the glandular epithelium. They were more numerous in the pyloric gland region than other locations of the alimentary tract. There are very few studies on the entero-endocrine cells of the reptiles. In one immunohistochemical study, immunoreactivity for somatostatin, gastrin, motilin, serotonin, pepsingen and bovine pancreatic polypeptide were detected in the mucosa of the alimentary tract of the King's skink (Arena *et al.*, 1990), and another one reported similar results in the gut of frogs (El-Salhy *et al.*, 1981). In different mammalian species such as rodents and feline parallel results were reported (Alumets *et al.*, 1977; Alumets *et al.*, 1979; Kitamura *et al.*, 1982a; Kitamura *et al.*, 1982b). These peptides and amines known to regulate the different digestive and motility functions of the alimentary tract (Krause *et al.*, 1985) and they are likely to do such in the *Varanus niloticus*. This first report on of the alimentary tract of the Egyptian *Varanus niloticus* highlighting the general morphology, mucous contents as well as locations and appearance of the endocrine cells may be the beginning of more interest in this species. Further studies are required to explore the histological and physio-logical characteristics of this animal species, to understand theory of development of reptilian species in comparison to known mammalian development, and to investigate the possibility of using this species in the experiments related to the digestive physiology.

## Acknowledgment

The authors thank Mr. Mohamed Abdullah and Mr. Elsayed Seedik, Faculty of Veterinary Medicine, South Valley University, for assistance in collecting the samples.

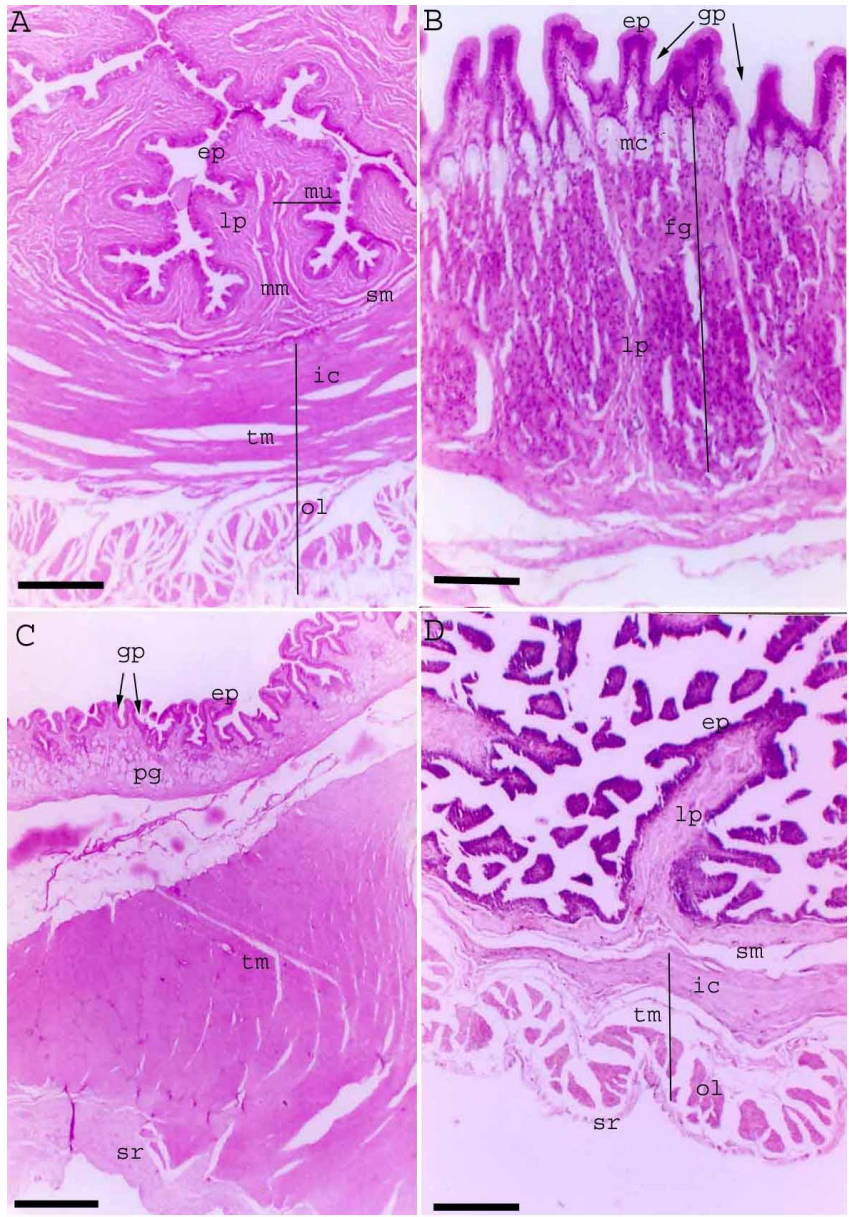

**Figure (1): General morphology of the alimentary tract of the *Varanus niloticus***
Paraffin sections from the esophageus (A), fundic (B) and pyloric stomach(C) , and
small intestine (D) stained with haematoxylin and eosin showing the general outline of
the alimentary tract of the Varanus Niloticus    . Surface lining epithelium (ep), gastric
pits (gp), lamina propria (lp), fundic glands (fg), pyloric glands (pg) muscularis mucosa
(mm), submucosa (sm), inner circular (ic) and outer longitudinal (ol) tunica muscularis
(tm) and serosa (sr).  Bars = 160 µm in parts A, C and D, and 40 µm in part B.

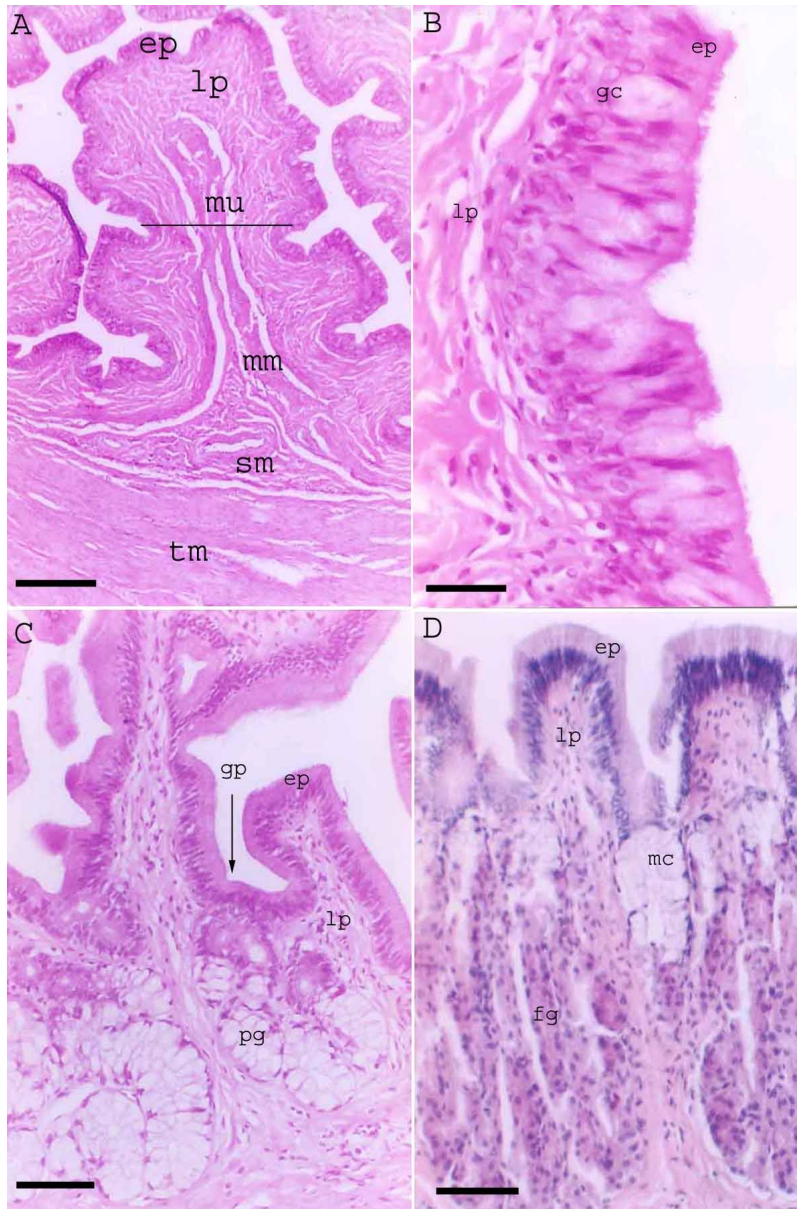

**Figure (2): Higher magnification showing details of some structures in the alimentary tract of the *Varanus niloticus.* Paraffin** sections from the esophageus (A, B), and fundic (C) and pyloric stomach stained with haematoxylin and eosin showing some details of the alimentary tract of the Varanus Niloticus . Surface lining epithelium (ep), mucous secreting goblet cells (gc), gastric pits (gp), lamina propria (lp), fundic glands (fg), mucous cells in fundic glands (mc), pyloric glands (pg), muscularis mucosa (mm), submucosa (sm) and tunica muscularis (tm).  Bars =   90 µm in A,   10 µm in B, and 20 µm in C and D.

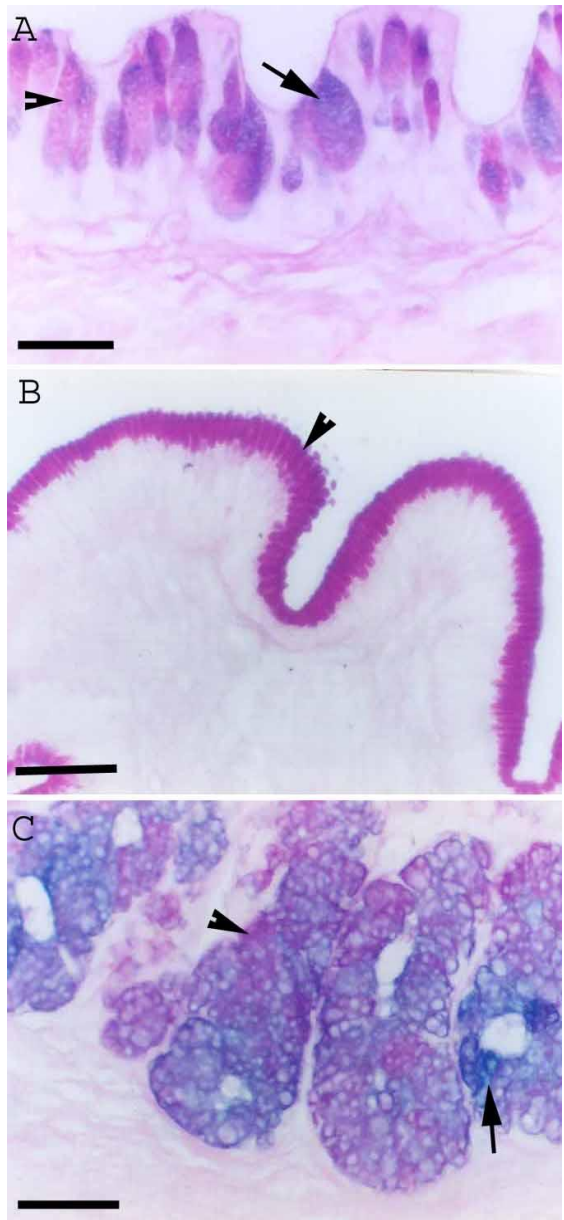

**Figure (3): Mucous contents of the alimentary tract of the *Varanus niloticus***

Paraffin sections from the surface lining epithelium of the esophageus (A), stomach (C) and pyloric mucous glands stained with PAS and AB. Arrowheads indicate PAS positive, and Arrows indicate AB positive reactions. Bars = 10

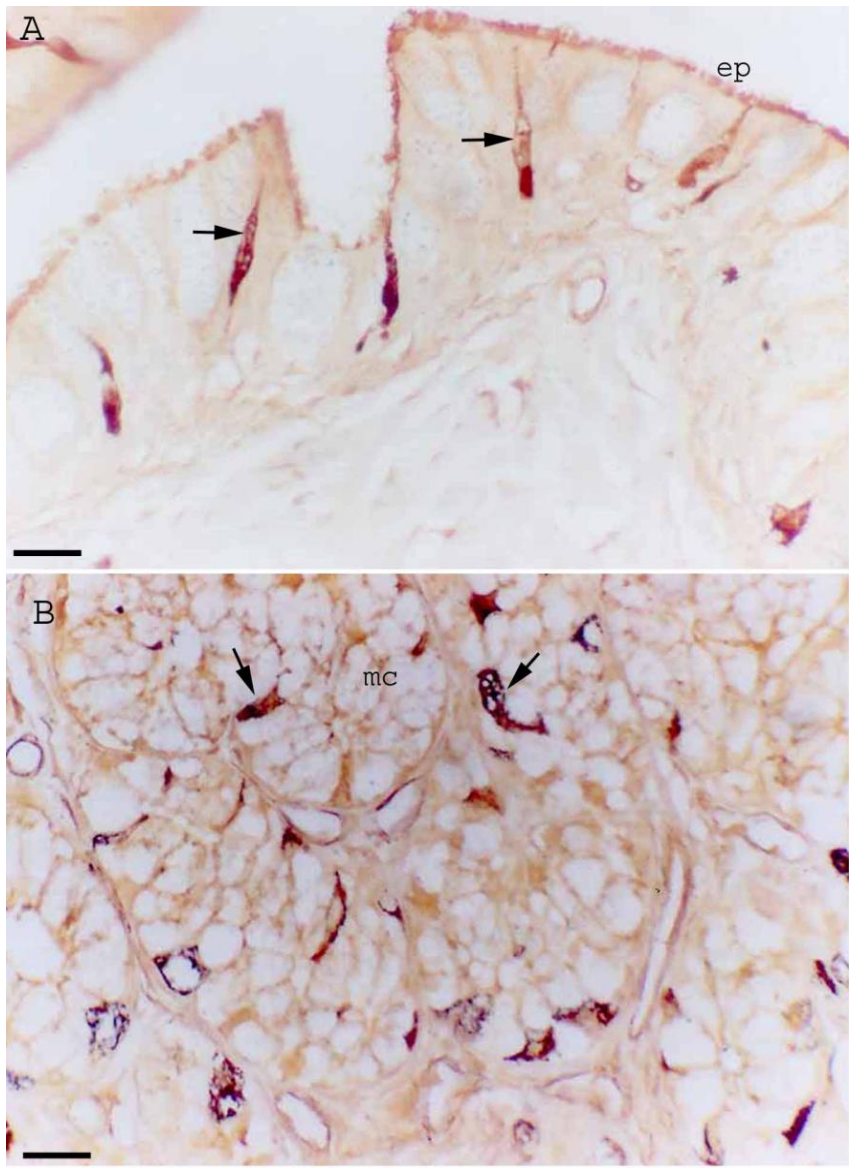

**Figure (4): Enteroendocrine cells localization in the alimentary tract of the *Varanus niloticus***

Paraffin sections from the lining epithelium of the esophageus (A) and the pyloric region of the stomach (B) stained according to Grimelius impregnation technique.  Arrows indicate enteroendocrine cells among the surface lining epithelium of the esophageus (A) and mucous cells of the pyloric glands (B).  Bars in parts A and B = 10